\documentclass[useAMS]{mn2e}
\usepackage{url}
\usepackage{graphicx}
\usepackage{soul}
\usepackage[T1]{fontenc}

\title[]{PF131010 Ciechan\'ow fireball - the body possible 
related to Near Earth Asteroids 2010 TB54 and 2010 SX11}
\author[A. Olech et al.]{A. Olech$^{1}$\thanks{e-mail:
olech@camk.edu.pl},  
P. \.Zo{\l}\k{a}dek$^2$, M. Wi\'sniewski$^{2,3}$, R. Rudawska$^4$, 
J. Laskowski$^2$, \newauthor K. Polakowski$^2$, M. Maciejewski$^2$, T. Krzy\.zanowski$^2$, T. Fajfer$^2$, and 
Z. Tymi\'nski$^5$\\
$^{1}$Nicolaus Copernicus Astronomical Center,
Polish Academy of Sciences, ul.~Bartycka~18, 00-716~Warszawa, Poland\\
$^{2}$Comets and Meteors Workshop, ul. Bartycka 18, 00-716 Warszawa, Poland\\
$^{3}$ Central Office of Measures, ul. Elektoralna 2, 00-139 Warsaw, Poland\\
$^{4}$ ESA European Space Research and Technology Centre, Noordwijk, The Netherlands\\
$^{5}$ Narodowe Centrum Bada\'n J\k{a}drowych, O\'srodek Radioizotop\'ow POLATOM, 
ul. So{\l}tana 7, 05-400 Otwock, Poland
}

\begin{document}

\date{Accepted 2015 March 15. Received 2015 February 20; in original form 2015 January 31}

\pagerange{\pageref{firstpage}--\pageref{lastpage}} \pubyear{2009}

\maketitle

\label{firstpage}

\begin{abstract}

On 2010 October 13, the Apollo type 20-meter asteroid 2010 TB54 passed
within 6.1 lunar distances from the Earth. On the same date, but 11.4
hours earlier, exactly at 02:52:32 UT, the sky over central Poland was
illuminated by $-8.6$ magnitude PF131010 Ciechan\'ow fireball. The
trajectory and orbit of the fireball was computed using multi-station
data of {\sl Polish Fireball Network (PFN)}. The results indicate that
the orbit of the meteoroid which caused the PF131010 fireball is
similar to the orbit of 2010 TB54 asteroid and both bodies may be
related. Moreover, two days before appearance of Ciechan\'ow fireball
another small asteroid denoted as 2010 SX11 passed close to the
Earth-Moon system. Its orbit is even more similar to the orbit of
Ciechan\'ow fireball parent body than in case of 2010 TB54.\\

The PF131010 Ciechan\'ow entered Earth's atmosphere with the velocity of
$12.8\pm0.2$ km/s and started to shine at height of $82.5 \pm 0.3$ km.
Clear deceleration started after first three seconds of flight, and the
terminal velocity of the meteor was only $5.8\pm0.2$ km/s at height of
$29.3 +\- 0.1$ km. Such a low value of terminal velocity indicates that
fragments with total mass of around 2 kg could survive the atmospheric
passage and cause fall of the meteorites. The predicted area of
possible meteorite impact is computed and it is located near Grabowo
village south of Ostro{\l}\k{e}ka city.

\end{abstract}


\section{Introduction}

On 15 February 2013 at about 03:20 UTC 17-meter asteroid entered the
Earth's atmosphere and exploded while travelling at a speed of 19 km/s.
The body became a superbolide meteor, which was seen over the southern
Ural region. At the moment of maximum brightness, which occurred near
the city of Chelyabinsk, the meteor was brighter than the Sun. The
several small fragments of meteorite (ordinary chondrite) were quickly
found on the west of Chelyabinsk, with the largest fragment of total
mass of 654 kg raised from the bottom of the Chebarkul lake on 16
October 2013 (Popova et al. 2013, Brown et al. 2013, Borovicka et al.
2013, Kohut et al. 2014).

About 16 hours after Chelyabinsk fireball occurrence the 45-meter
asteroid 367943 Duende (2012 DA14) approached the Earth and missed it by
about 27 700 km. This was striking coincidence but subsequent analysis
of the data clearly indicated the two objects could not have been
related because they had widely different orbits (Wlodarczyk 2012,
Moskovitz et al. 2013)

What is more interesting similar situation occurred over two years
earlier. On the night of 2010 Oct 9/10, Mt. Lemmon Survey reported a
discovery of a new 20-meter Apollo type asteroid designated as 2010 TB54
(Hergenrother 2010). The asteroid has been expected to pass only 6.1
lunar distance (0.011 AU) from the Earth. Over 11 hours earlier sky over
central Poland was illuminated by $-8.6$ magnitude fireball. What is
even more interesting these both bodies, contrary to Chelyabinsk
meteorite parent body and Duende asteroid, seem to be related.

In this paper we report an analysis of the multi-station observations
of the PF131010 fireball made by cameras of the {\sl Polish Fireball Network}.
Both the trajectory and orbit are calculated indicating that the fireball
was related to 2010 TB54 asteroid. Even closer resemblance of orbits is
found for 2010 SX11 asteroid.

\section{Observations}

The {\sl Polish Fireball Network (PFN)} is the project whose main goal
is regularly monitoring the sky over Poland in order to detect bright
fireballs occurring over the whole territory of the country (Olech et
al. 2005, \.Zo{\l}\k{a}dek et al. 2007, 2009, Wi\'sniewski et al. 2012).
It is kept by amateur astronomers associated in {\sl Comets and Meteors
Workshop (CMW)} and coordinated by astronomers from Copernicus
Astronomical Center in Warsaw, Poland. Currently, there are over 20
fireball stations belonging to {\sl PFN} which operate during each clear
night.

\begin{figure}
\centering
\includegraphics{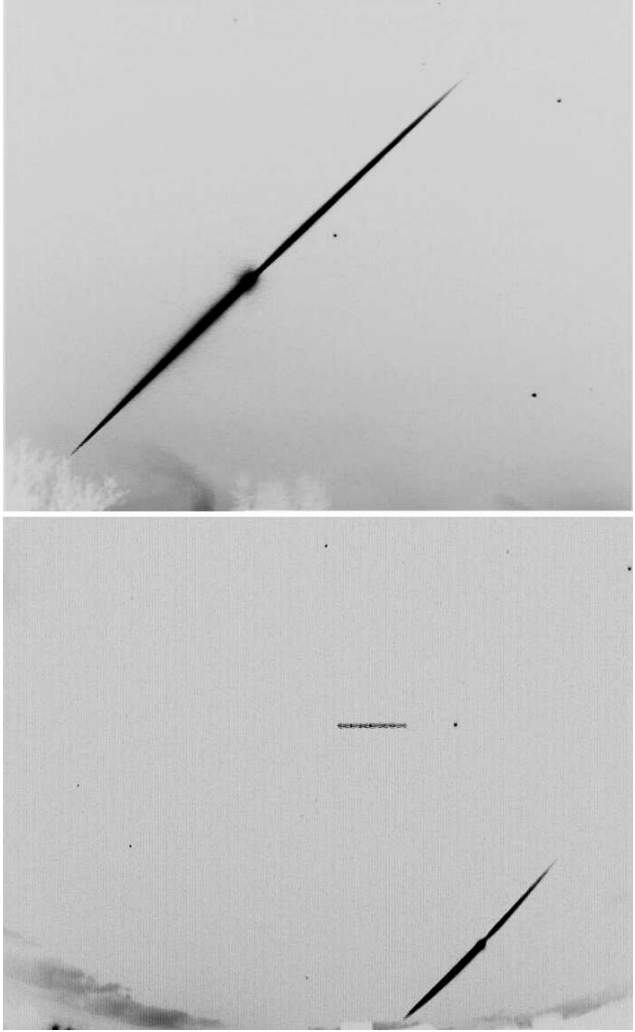}
\vspace{14.3cm}
\caption{The video images of PF131010 Ciechan\'ow fireball captured in
Nowe Miasto Lubawskie (upper panel) and Gniewowo (lower panel).}
\end{figure}

The PF131010 Ciechan\'ow fireball was observed by five {\sl PFN} video
stations (Table 1): two of the recordings allowed us to determine the
trajectory of the phenomenon, one recording includes  observations made
from a distance of over 400 km, another one features an initial part of
the trajectory, and the last one noticed a bright reflection on clouds
layer only.

The most detailed and valuable recording comes from the PFN37 station in
Nowe Miasto Lubawskie, positioned 80 km north-east of Toru\'n (see Fig.
1). A camera with a 70-degree field of view  registered its path in the
southern direction diagonally through the field of view. The bolide
appeared between the constellations of Orion and the Gemini as a
point-like structure of about 2 magnitude. The brightness of the meteor
increased quickly, and after 1.2 seconds the wake appeared. Until about
3 seconds the bolide continued its luminous path as an object of about
$-4$ magnitude. The fireball started to lose its brightness in the
middle  of the visible trajectory, reaching a local minimum at about 4.6
second of the flight.

\begin{table*}
\centering
\caption[]{Basic data on the PFN stations which recorded PF131010 Ciechan\'ow fireball.}
\begin{tabular}{|l|l|c|c|c|l|l|}
\hline
\hline
Code & Site & Longitude [$^\circ$] & Latitude [$^\circ$] & Elev. [m] & Camera & Lens \\
\hline
PFN13 & Toru\'n & 18.6209 E & 53.0252 N & 65 & Siemens CCBB1320-MC & Ernitec 4 mm f/1.2  \\
PFN24 & Gniewowo & 18.3042 E & 54.5779 N & 130 & Siemens CCBB1320-MC & Ernitec 4 mm f/1.2  \\
PFN32 & Che{\l}m  & 23.4982 E & 51.1355 N & 194 & Mintron 12V8HC-EX & VM2312 2.3-6 mm \@ 4 mm \\
PFN37 & Nowe Miasto Lubawskie  & 19.5922 E & 53.4349 N & 95 & Tayama C3102-01A1 & Computar 4 mm f/1.2 \\
PFN38 & Podg\'orzyn & 15.6817 E & 50.8328 N & 360 & Tayama C3102-01A4 & Evatar 3.5-8 mm \@ 4 mm  \\
\hline
\hline
\end{tabular}
\end{table*}

Just after the minimum, a sudden increase of brightness to about $-8.5$
magnitude was registered. It means that the bolide most probably fell
apart. However, on the basis of the video data, it would be difficult to
describe the whole process in more details. The image of the moving
fireball is strongly overexposed at that moment, one can notice only an
elongated structure following the bolide and disappearing after less
than 0.2 of a second. From that point the brightness of the bolide
decreases constantly and evenly, without any visible fluctuations. After
over 9.5 seconds of the  luminous trajectory of the bolide, becoming an
object with a brightness  lower than 0 magnitude, hiding behind a bough
of a tree visible near the horizon and could not be seen again on the
other side of that obstacle.

The second recording, pertinent to the analysis of that phenomenon, was
obtained from the PFN24 Gniewowo station, situated 15 km north-west of
Gdynia (see the second panel of Fig. 1). Because of the station's
technical problems we got just a static image, without any possibility
of recovering data from video frames. However, a file with the XY
coordinates was saved along with it, and on that basis one can
reconstruct the position of the bolide. The image of that phenomenon is
similar to that, registered by the Nowe Miasto Lubawskie station but, in
this case, the distance is notably larger; its position is much closer
to the horizon with lower elevation angles.

The third of the stations -- PFN38 Podg\'orzyn -- was situated on
the opposite side of Poland, 10 km of Jelenia G\'ora. From a distance of
several hundred kilometers the phenomenon was visible just over the
horizon. It is interesting that there is a persistent train which
does not disappear for about 0.5 of a second in the place of the flare.
That train can be connected to the elongated structure, visible after the
flare in the Ciechan\'ow station images. The recordings of the Podg\'orzyn
station show that the bolide ended right over the horizon, on the edge
of the camera's field of view. 

A very small fragment of the fireball was registered by the PFN32 Che{\l}m
station. The PFN13 Toru\'n station managed to notice a distinct bright
reflection on a thin clouds layer covering the sky. The fireball was not
observed by other fireball networks in central Europe.

On the basis of all records, we determined the momentum of the
phenomenon. The moment of maximum brightness was at 02:52:32 UT $\pm$ 2
seconds. The beginning of the phenomenon observed by PFN37 station was
recorded at 02:52:28 UT, while the last point of the bolide was visible
at 02:52:38 UT.

The observations were made using CCTV cameras with comparable
parameters. All the cameras worked in PAL resolution $786\times584$ 
with 25 frames per second offering 0.04 second temporal resolution. The
monochrome CCD detectors equipped with fast CCTV lenses of focal length
of around 4 mm were used. A typical limiting magnitude of that kind of
equipment amounts to $+2$. The details concerning coordinates all the
stations and equipment used are summarized in Table 1.

\section{Data reduction}

Using the data gathered by the {\sl PFN} network cameras an analysis of
the phenomenon was conducted. Because of the huge differences in
distance from the fireball and, in consequence the quality of the
images, only two best recordings were taken into account: from the PFN37
Nowe Miasto Lubawskie station and from the PFN24 Gniewowo station, as
both videos show the most complete flight of the bolide. The data from
these stations, after a previous conversion, were further reduced
astrometricaly by the {\sc UFO Analyzer} program (SonotaCo 2009).
Initially only automatic data were taken into account but during the
further processing it became obvious that significant overexposures, the
presence of the wake and a possible fragmentation after the flare caused
quite serious errors concerning the correct position of the points of
the phenomenon. The measurement precision improved noticeably when the
bolide's position was determined using UFO Analyzer astrometric solution
with  manual centroid measurement {\sc UFO Analyzer}.

The position of the meteor was recognized on 241 frames of the video
from PFN37 station, which means that the time of the flight lasted 9.64
seconds. The photometry of the phenomenon was done with the help of the
{\sc IRIS} program, using reference stars, Jupiter and the Moon at phase
close to New Moon as comparison objects. The trajectory of the
phenomenon was determined using the {\sc PyFN} software
(\.Zo{\l}\k{a}dek 2012).

\section{Results}

\subsection{Trajectory of the fireball}

The Ciechan\'ow fireball moved from west to east following moderately steep trajectory. 
The trajectory azimuth and the trajectory slope were 261
degrees and 29 degrees, respectively. The beginning of the
bolide was situated in a place with the following coordinates:
$\phi=52.831(2)^\circ$~N, $\lambda=19.901(1)^\circ$~E at the height of
$82.5\pm0.3$ km. That location is 33 km north of P{\l}ock. In next
seconds the bolide traveled east and flew 2 kilometers north of
Ciechan\'ow, reaching its maximum brightness at the height of $54.4\pm0.1$
km. Then it went 10 km south of Przasnysz, 9 km north of Mak\'ow
Mazowiecki and it reached its ending point 10 km north-west of
R\'o{\.z}an. It was situated then at the height of $29.3\pm0.1$ km over
the following coordinates: $\phi=52.962(2)^\circ$~N, 
$\lambda=21.275(5)^\circ$~E. The trajectory of the PF131010 fireball is
shown in Fig. 2 and all important parameters are summarized in Table 2.

\begin{figure}
\centering
\includegraphics{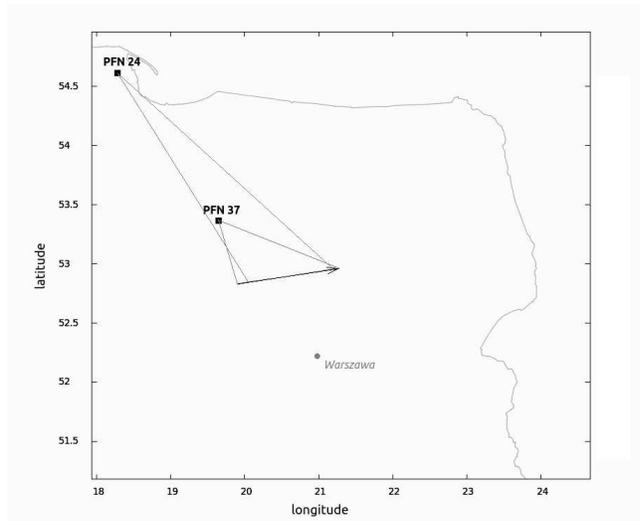}
\vspace{7.8cm}
\caption{The luminous trajectory of the PF131010 fireball over eastern Poland 
and the location of the PFN stations which data were used in calculations.}
\end{figure}  

\begin{table*}
\centering
\caption{Characteristics of the PF131010 Ciechan\'ow fireball}
\begin{tabular}{lccc}
\hline
\multicolumn{4}{c}{2010 October 13, ${\rm T} = 02^h52^m32^s \pm 2.0^s$ UT}\\
\hline
\multicolumn{4}{c}{Atmospheric trajectory data}\\
\hline
 & {\bf Beginning} & {\bf Max. light} & {\bf Terminal} \\
Vel. [km/s] & $12.9\pm0.2$ & $12.7\pm0.2$ & $5.8\pm0.2$ \\
Height [km] & $82.5\pm0.3$ & $54.4\pm0.1$ & $29.3\pm0.1$ \\
Long. [$^\circ$E] & $19.901\pm0.001$ & $20.622\pm0.002$ & $21.275\pm0.005$\\
Lat. [$^\circ$N] & $52.831\pm0.002$ & $52.902\pm0.001$ & $52.962\pm0.002$\\
Abs. magn. & $-1.6\pm0.5$ & $-8.6\pm0.5$ & $1.5\pm1.0$ \\
Slope [$^\circ$] & $29.2\pm0.1$ & $28.8\pm0.1$ & $28.4\pm0.1$\\
Duration & \multicolumn{3}{c}{9.4 sec}\\
Lenght & \multicolumn{3}{c}{$110.5\pm0.8$ km}\\
Stations & \multicolumn{3}{c}{Nowe Miasto Lubawskie, Gniewowo, Podg\'orzyn, Che{l}m, Toru\'n}\\
\hline
\multicolumn{4}{c}{Radiant data (J2000.0)}\\
\hline
 & {\bf Observed} & {\bf Geocentric} & {\bf Heliocentric} \\
RA [$^\circ$] & $19.93\pm0.24$ & $6.17\pm0.36$ & - \\
Decl. [$^\circ$] & $17.68\pm0.04$ & $0.41\pm0.42$ & - \\
Vel. [km/s] & $12.9\pm0.2$ & $6.9\pm0.3$ & $32.2\pm0.3$\\
\hline
\end{tabular}
\end{table*}

\subsection{Velocity}

Based on those observations the velocity of the phenomenon was
determined for different points of its trajectory. In the initial part
of the trajectory the velocity did not change in a noticeable way. The
initial velocity of the bolide was close to the lower limit of
meteoroids entering the atmosphere of the Earth and amounted to
$12.8\pm0.2$ km/s.  After the third second of the flight the velocity
decrease became clearly visible, by $t=$ 6.5 seconds amounting to 12
km/s; from about 7 seconds the  intensive slowing down of the residue of
the meteoroid with more or less constant value of about 2700~${\rm
m/s^2}$ can be noticed. The fireball stopped being visible (disappearing
behind trees near the horizon) at the velocity of $5.8\pm0.2$ km/s. The
time of its further flight was very short, most possibly far below 1
second; otherwise the phenomenon would have been observed through a
crack between trees on the extension line of the phenomenon visible
part. The evolution of the velocity of PF131010 Ciechan\'ow fireball is
shown in Fig. 3.

\begin{figure}
\centering
\includegraphics{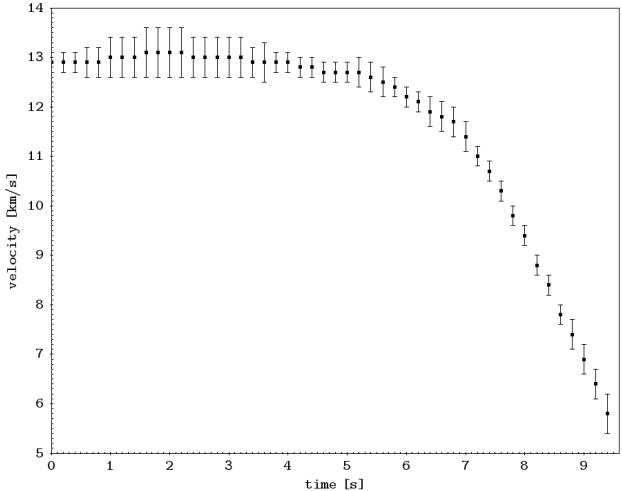}
\vspace{8cm}
\caption{The evolution of the velocity of the PF131010 Ciechan\'ow fireball.
Initial velocity given by observations was assumed as a preatmospheric velocity,
initial velocity was nor fitted due to limited data precision. Error bars has been
determined by repeating calculations with various input values according to astrometry errors.}
\end{figure}

\subsection{Brightness}

The light curve of the Ciechan\'ow fireball is shown in Fig. 4 and has
some interesting features. It should be emphasized especially the fact
that there are two separate parts of the curve, distinctly differing in
luminosity. Between its initial phase and the final phase you can notice
a flare of a maximum absolute magnitude amounting to $-8.6\pm0.5$.

Both parts of the curve themselves have gentle characteristics. In the
initial phase there were no brightness oscillations. The luminosity
increases from the border of the camera limit to about $-4$ magnitude in
about 1.5 seconds. The further increase is a bit slower, after 3.5
seconds the bolide reaches a local maximum amounting to $-5.8$
magnitude. For the next second the brightness of the bolide decreases to
a value of $-4$ magnitude. By t = 4.5 seconds one can see a sudden
increase of brightness when the luminosity increases by five magnitudes
in several tenths of a second. The flare takes place at 55 km with the
dynamic pressure being rather low and amounting to 0.05 Mpa. The dynamic
pressure has been calculated using formula $p= \Gamma\rho {v}^{2}$,
using $\Gamma = 0.921$ (Carter et al. 2009) and  air density given by
MSISE-90 atmospheric model (calculated for the exact moment of the
event).

\begin{figure}
\centering
\includegraphics{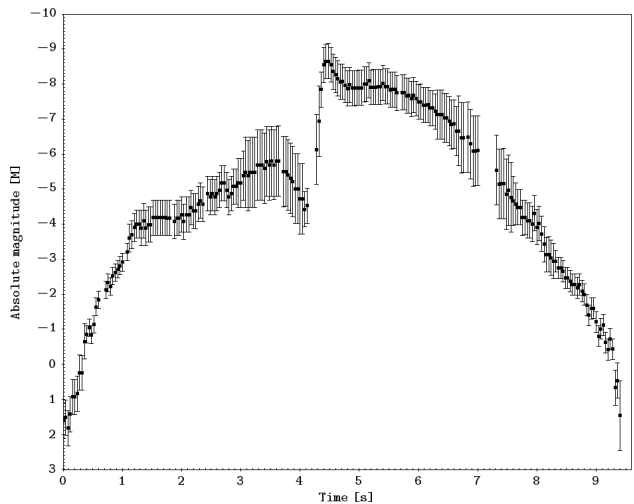}
\vspace{8cm}
\caption{The light curve of the PF131010 Ciechan\'ow fireball. Brightness has been
determined using aperture photometry with bright stars and the Moon images registered
by the same camera used as a reference objects. Errors has been determined by repeating
measurements with various reference objects and settings.}
\end{figure}

The second part of the curve starts with a brightness increase; then the
luminosity decreases slowly during next seconds but keeping a very high
level, reaching the value before the flare only in the next 3 seconds.
For the decreasing part of the light curve some luminosity changes of
0.5 magnitude can be seen at $t=7.8$ second and the dynamic pressure of
0.35 Mpa. A small fragmentation in that place is possible. In the final
phase of the phenomenon the curve is very gentle, the fireball is fading
down slowly, reaching a brightness of $+2$ magnitude in the last frame.
A light curve with such properties as the ones shown above might
indicate some interesting properties of the meteoroid. Observed flare
and brighter part of lightcurve may by caused by large number of small
fragments separating from the main body after $t=$ 4.5 second. There were no
clearly visible fragments on any video record but bright feature visible
on the video from PFN38 camera may suggest that the large number of
small fragments ablated shortly after $t=$ 4.5 second. Further meteoroid
erosion may be responsible for the brightness increase in second part of
the trajectory. Such mechanism of meteoroid erosion has been recently
modeled for Kosice fireball (Borovi\v{c}ka et al. 2013). Another
possibility  is a distinct change of the properties of material
undergoing ablation. The meteoroid may be covered with some kind of a
shell with lower ablation coefficient.

\begin{table*}
\caption[]{Orbital elements of the PF131010 fireball compared to the orbits of
2010 TB54 and 2010 SX11 asteroids.}
\centering
\begin{tabular}{|l|c|c|c|c|c|c|c|c|}
\hline
\hline
  & $1/a$ & $e$ & $q$ & $\omega$ & $\Omega$ & $i$ & P & $D'$ \\
  & [1/AU] & & [AU] & [deg] & [deg] & [deg] & year & \\ 
\hline
PF131010  & 0.837(3)  & 0.263(3)  &  0.880(3) & 64.14(82) & 19.5248(1) & 0.45(5) & 1.304(10) & \\
2010 TB54 & 0.8680(3) & 0.2791(3) & 0.8308(2) & 82.566(4) & 14.2330(3) & 6.62(1) & 1.2366(7) & 0.058\\
2010 SX11 & 0.8703(3) & 0.2495(2) & 0.8703(3) &266.873(4) & 182.524(1) & 5.17(1) & 1.2487(5) & 0.043\\
\hline
\hline
\end{tabular}
\end{table*}  

\subsection{Dark flight and a possible fall of the meteorite}

The observed final velocity of 5.8 km/s at the height of 29 km (which
most probably could descend even lower) indicates a possible meteorite
fall near Grabowo, south of Osto{\l}\k{e}ka. An estimated impact point
of the single meteorite weighing about 2 kilograms is about 22
kilometers of the final trajectory point, about 3.5~km to the left of
the trajectory axis (see Fig. 5). The exact coordinates of the impact
point are $\phi=52.961(5)^\circ$~N, $\lambda=21.624(30)^\circ$~E.
Calculations have been performed with assumption that only one fragment
survived the ablation.  The dynamic mass has been calculated using
parameters observed at the end of the visible trajectory. The mass is
given for chondrite body with bulk density of 3.7 ${\rm g}\cdot {\rm
cm}^{3}$ and calculated  using standard formulas (Ceplecha 1987). 

\begin{figure}
\centering
\includegraphics{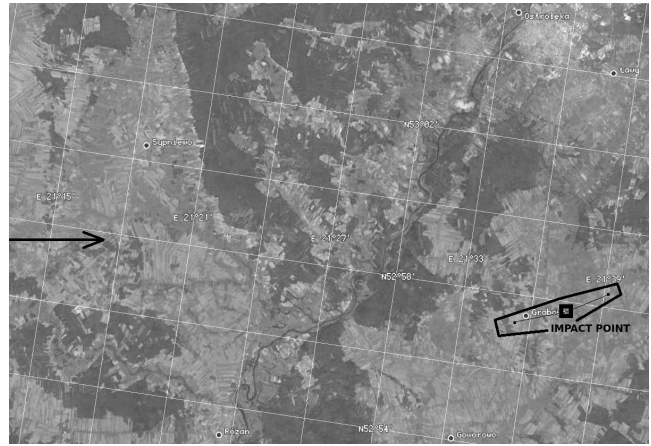}
\vspace{6.7cm}
\caption{Map with computed impact area of meteorite caused by the PF131010 Ciechan\'ow fireball.
The final part of the trajectory is also shown.}
\end{figure}

\begin{figure}
\centering
\includegraphics{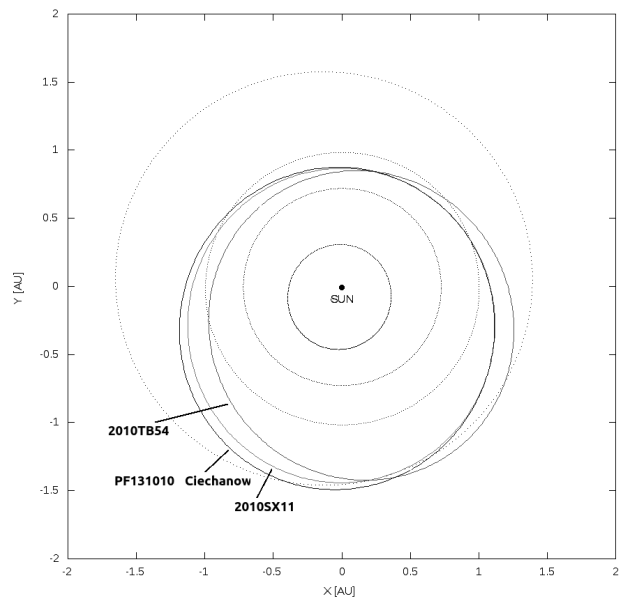}
\vspace{8cm}
\caption{Plot of the inner Solar System with orbits of the PF131010 Ciechan\'ow fireball
and two asteroids: 2010 TB54 and 2010 SX11.}
\end{figure}

Because of huge trajectory uncertainties concerning the parameters
observed in the final part of the trajectory the precision in
determining the final place of impact amounts to about 3 km. 

An atmosphere profile, obtained during atmospheric probing performed on
October 13, 2010 at 00:00 UT (the Legionowo meteorological station located 65~km 
south of the terminal point), was used for the computations. 

A searching expedition was launched soon after the first rough estimates
of results was determined, but their efforts were fruitless. After
another manual measurement of the meteor position there was a new result
(presented here), slightly different from the previous one. The second
expedition took place at the end of March 2015, though without positive
results either. The area (with active agriculture) is not easy to
search, with at least half of it not easily accessible.

\subsection{Orbit}

Based on the observational data we were able to determine the radiant of
Ciechan\'ow fireball, its geocentric velocity and orbital  parameters of
the meteoroid which entered the Earth's atmosphere. The orbital
parameters are listed in the first row of Table 3\textbf{,} and the
diagram showing orbit of the fireball in the inner Solar System is
displayed in Fig. 6.

The orbit of Ciechan\'ow fireball is located almost in the ecliptic
plane and has low eccentricity. The meteoroid hit the Earth less than
two months before perihelion passage which was expected on December 4,
2010 at distance of $q=0.880\pm0.003$ AU. 

Comparison of the orbit of Ciechan\'ow fireball  to  orbits of Near
Earth Objects (NEO) allowed us to select several asteroids with Drummond
criterion $D_D<0.109$ (Drummond 1979). Two of them are especially
interesting. The 2010 TB54 asteroid has $D_D$ criterion value as small
as 0.058. What is more interesting this object passed within the
distance of only 0.016 AU from the Earth on October 13, 2010 at 14:14
UT, i.e. only 11.4 hours before the occurrence of the Ciechan\'ow
fireball. This asteroid has been discovered by Mount Lemnon Survey on
October 9, 2010, its observing arc is 3 day long and based on 30 optical
measurements. 2010~TB54 has absolute magnitude $+26.8$ magnitude, and
its diameter is less than 29 meters.

Even lower Drummond criterion value ($D_D=0.043$) was noted for 2010
SX11 asteroid. The close encounter with this body occurred on October
11, 2010 at 13:10 UT at the distance of 0.025 AU. In this case the time
difference amounts to 37.7 hours. 2010 SX11 is a larger body, with
absolute magnitude $+24.8$ magnitude and diameter between 33 and 73
meters. Observing arc span is 21 days, 33 optical measurements were used
to determine orbital elements. Uncertainties of orbital elements for
both asteroids are similar but slightly smaller for 2010 SX11.

Our results indicate the possibility that in period of October 11-13
there is an activity of meteor shower of asteroid origin with radiant 
located in the border of Pisces and Aries constellations.

\section{Modeling}

A numerical integration of the orbital parameters backwards in time has
been performed in order to test the link between the fireball
Ciechan\'{o}w and two NEOs: 2010 SX11 and 2010 TB54. For the
integrations of the asteroids and test particles representing fireball,
the RADAU integrator in the Mercury software was
used (Chambers 1999). The model of the Solar System used in
integrations included: 8 planets, four asteroids (Ceres, Pallas, Vesta,
and Hygiea), and the Moon as a separate body. The positions and
velocities of the perturbing planets and the Moon were taken from the
DE406 (Standish 1998). The initial orbital elements of asteroids
2010 SX11 and 2010 TB54 were taken from JPL HORIZONS
website~\footnote{http://ssd.jpl.nasa.gov/?horizons}. Together
with initial orbital elements of asteroids, the test particles were
integrated to the same epoch of the beginning of the integration. Next,
the backward integration was continued for 5000 yr.

During the evolution, the ascending and descending nodes of theoretical
particles are dispersed within heliocentric distances from 0.8 to 1.8
AU, with a concentration around Earth's orbit. The generated stream has
been widely dispersed in longitude, mostly by perturbation from the
Earth. Therefore, application of a conventional similarity functions:
$D_{SH}$ (Southworth and Hawkins 1963), $D_D$ (Drummond 1981), or $D_J$
(Jopek 1993), would be strongly influenced in the longitude term in the
D-criterion. Due to it, we used (Steel 1991) criterion, $D_S$, where the
longitude term is not included. Figures 7 and 8 show that the evolution
of the $D_S$ criterion reveals a link between the Ciechan\'{o}w fireball
and NEOs, with the values of $D_S$ being less than 0.15 through the
whole integration time (except one test particle when we compare orbits
with 2010 TB54 and 2010 SX11, respectively).

The theoretical geocentric radiants of the asteroids has been determined
using Fortran code which is able to calculate radiant coordinates and 
theoretical stream orbit (Neslusan et al. 1998). Theoretical radiants of
asteroids and Ciechan\'{o}w meteoroid doesn't match if calculated from
present orbital elements. However similarity of the radiant is visible
for back integrated orbital elements. Distance between 2010~SX11 radiant
and Ciechan\'{o}w radiant was close to 8.5 degrees 5000 years ago. Also
the theoretical radiant of 2010~TB54 was located in the same sky area,
7.5 degrees from the Ciechan\'{o}w radiant. Five thousands years ago all
theoretical radiants were closer than 10 degrees each other with
geocentric discriminants $D_D=0.044$ for 2010~SX11, $D_D=0.087$ for
2010TB54 and $D_D=0.010$ for Ciechan\'{o}w. Change of radiant distances
in time may suggest that age of possible stream is probably larger than
10000 years.

\begin{figure}
\centering
\includegraphics{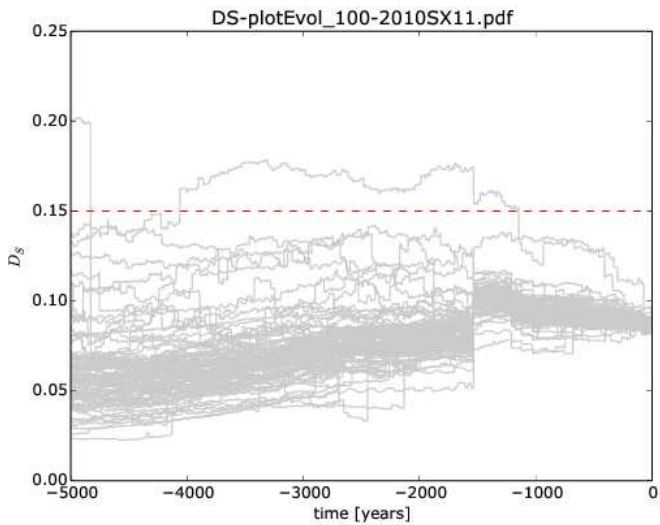}
\vspace{7.0cm}
\caption{Evolution of the $D_S$ criterion calculated by comparing the 
orbit of 2010~SX11 and those of test particles. The red dashed line 
shows the threshold value ($D_c=$0.15)}
\end{figure}

\begin{figure}
\centering
\includegraphics{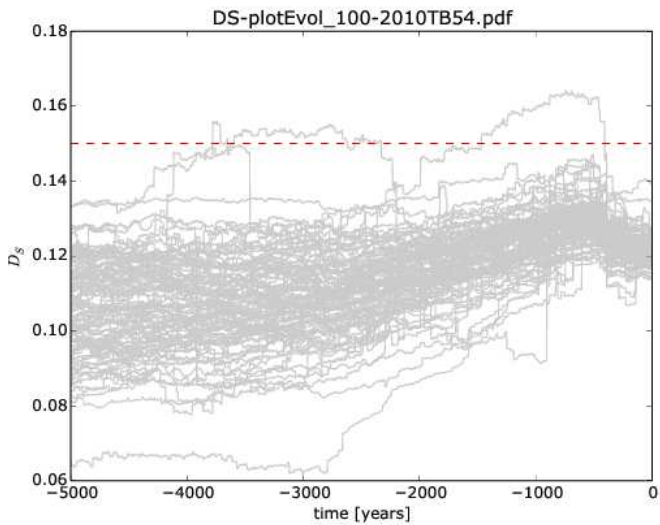} 
\vspace{7.0cm}
\caption{Evolution of the $D_S$ criterion calculated by comparing 
the orbit of 2010~TB54 and those of test particles. The red dashed 
line shows the threshold value ($D_c=$0.15)}
\end{figure}

\section{Summary}

In this paper we presented an analysis of the multi-station
observations of a bright fireball which occurred over eastern
Poland. Our main conclusions are as follows:

\begin{itemize}

\item the meteor appeared on 2010 Oct 12/13 at 02:52:32 UT over
the eastern part of Poland  was detected by five video 
stations of {\sl Polish Fireball Network},

\item the maximum brightness of the fireball reached $-8.6\pm0.5$ mag
and was observed at height of $54.4\pm0.1$ km over Ciechan\'ow city,

\item the entry velocity was only $12.8\pm0.2$ km/s and after three
seconds of flight the meteoroid was significantly decelerated with
the rate of $2700~ {\rm m}\cdot {\rm s}^{-2}$ resulting with final
velocity of only $5.8\pm0.2$ km/s,

\item low value of final velocity indicates a possible 2 kg meteorite
fall near Grabowo, south of Osto{\l}\k{e}ka, 

\item the low eccentric orbit of the fireball, positioned almost in the
ecliptic plane,  is similar to orbits of 2010~TB54 and 2010~SX11
asteroids, which passed the Earth 11.4 and 37.7 hours before the
occurrence of the fireball, respectively,

\item numerical integration of the orbital elements backwards in time
indicates that Ciechan\'ow fireball and 2010~TB54 and 2010~SX11
asteroids may have common origin.

\end{itemize}

\section*{Acknowledgments}

This work was supported by the NCN grant number 2013/09/B/ST9/02168.

\end{document}